\documentclass[preprint2, flushrt]{aastex}
\newcommand{\re}{\par\hangindent=0.5cm\hangafter=1\noindent}

\slugcomment{To appear in {\it Astrophysical Journal Letters}}
\shorttitle{Black Hole-to-Bulge Relations}
\shortauthors{Umemura}

\begin{document}

\title{A Radiation-Hydrodynamical Model for Supermassive 
Black Hole-to-Bulge Mass Relation and Quasar Formation
}

\author{Masayuki Umemura}
\email{umemura@rccp.tsukuba.ac.jp}
\affil{Center for Computational Physics, University of Tsukuba,
Tsukuba, Ibaraki 305, Japan\\[3mm]
{\rm To appear in} {\it Astrophysical Journal Letters}
}

\begin{abstract}
As a potential mechanism to build up supermassive black holes (BHs)
in a spheroidal system, 
we consider the radiation drag effect by bulge stars, which extracts angular
momentum from interstellar gas and thus allows the gas to accrete onto the galactic 
center. With incorporating radiation hydrodynamical equation with
simple stellar evolution, it is shown that the BH-to-bulge mass ratio, 
$f_{\rm BH}$, is basically determined by a fundamental constant, that is,
the energy conversion efficiency for nuclear fusion of hydrogen to helium, 
$\varepsilon=0.007$. More specifically, $f_{\rm BH}$ is
predicted to be $0.3\varepsilon -0.5\varepsilon$.
Based on the present model for BH growth, a scenario for quasar formation is 
addressed in relation to ultraluminous infrared galaxies.
\end{abstract}


\section{Introduction}

The recent compilation of the kinematical
data for galactic centers in both inactive and active galaxies
has revealed that the estimated mass of a central 'massive dark object' (MDO), 
which is the nomenclature for a supermassive BH candidate,
does correlate with the properties of galactic bulges.
The demography of MDOs has shown a number of intriguing relations:
1) The BH mass exhibits a linear relation to the bulge mass with
the ratio of $f_{\rm BH}\equiv M_{\rm BH}/M_{\rm bulge}=0.001-0.006$
as a median value (Kormendy \& Richstone 1995; Magorrian et al. 1998; 
Richstone et al. 1998;
Gebhardt et al. 2000; Ferrarese \& Merritt 2000; Merritt \& Ferrarese 2001a)
 (It is noted that the bulge means a whole galaxy for an elliptical galaxy.)
2) The BH mass correlates with the velocity dispersion of bulge
stars with a power-law relation as $M_{\rm BH} \propto \sigma^n$,
$n=3.75$ (Gebhardt et al. 2000a) or 4.72 
(Ferrarese \& Merritt 2000; Merritt \& Ferrarese 2001a, 2001b). 
3) The ratio $f_{\rm BH}$ tends to grow with the age of youngest stars 
in a bulge until $10^9$yr (Merrifield et al. 2000).
4) In disk galaxies, $f_{\rm BH}$ is significantly smaller than 0.001
when the disk stars are included (Salucci et al. 2000; Sarzi et al. 2001). 
5) For quasars, $f_{\rm BH}$ is level with
that for elliptical galaxies (Laor 1998).
6) As for Seyfert galaxies, the mass ratio is under controversy.
The estimation by the reverberation mapping suggests that 
$f_{\rm BH}$ in Seyfert 1 galaxies is considerably smaller 
than 0.001 (Wandel 1999; Gebhardt et al. 2000b),
although the reverberation mapping method 
may cause a systematic underestimate of BH mass (Krolik 2001).
However, the measurements of H$_\beta$ line widths give
$f_{\rm BH}\simeq 0.0025$ for Seyfert 1's as well as
quasars (McLure \& Dunlop 2001).
On the other hand, the BH mass-to-velocity dispersion relation 
in Seyfert 1 galaxies seems to hold good in a similar way to elliptical 
galaxies (Nelson 2000; Gebhardt et al. 2000b).
All of these correlations imply that
the formation of a supermassive BH is likely to be physically
linked to the formation of a galactic bulge which harbors
a supermassive BH. 
In addition, a great deal of recent efforts have revealed 
that quasar host galaxies 
are mostly luminous and well evolved early-type galaxies
(McLeod \& Rieke 1995; Bahcall et al. 1997; Hooper, Impey, \& Foltz 1997;
McLeod, Rieke, \& Storrie-Lombardi 1999; 
Brotherton et al. 1999; Kirhakos et al. 1999;
McLure et al. 1999; McLure, Dunlop, \& Kukula 2000). 
These findings, combined with the BH-to-bulge relations, suggest
that the formation of a supermassive BH, a bulge, and a quasar is 
mutually related.

Some theoretical models have been considered to explain the
BH-to-bulge correlations,
e.g., hydrodynamical ones including a wind-regulation 
model (Silk \& Rees 1998) 
or an inside-out accretion model (Adams, Graff, \& Richstone 2001), 
and self-interacting dark matter model (Ostriker 2000).
But, little has been elucidated regarding the physics on
the angular momentum transfer which is inevitable for BH formation,
since the rotation barrier by the tidal spin up 
in a growing density fluctuation is given by
$$
{R_{\rm barr} \over R_{\rm Sch}}
\approx 10^6 \left({M_b \over 10^{10} M_\odot}\right)^{-2/3}
\left({\lambda \over 0.05}\right)^2(1+z)^{-1}
$$
in units of the Schwarzshild radius $R_{\rm Sch}$,
where $M_b$ is the baryonic mass, $z$ is the cosmological redshift,
and $\lambda$ is the spin parameter which
provides the ratio of circular velocity to velocity dispersion of dark matter
(Sasaki \& Umemura 1996). Here, $R_{barr}$ is estimated by
$R_{barr}=j_b(z)^2/GM_b$, where $j_b$ is the specific angular momentum
given by $j_b \simeq R_{\rm max}\sigma\lambda$ 
with $R_{\rm max}$ being the maximum expansion radius of the fluctuation. 
Furthermore, required mechanisms for BH formation must work
effectively in a spheroidal system like a bulge.
The $\alpha$-viscosity or non-axisymmetric gravitational instabilities 
would effectively transfer angular momentum 
once a disk-like system forms, but they are not likely to work
in a spheroidal system.
In this {\it Letter}, 
as a potential mechanism in a spheroidal system, the relativistic drag
force by the radiation from bulge stars is considered,
and the BH-to-bulge ratio is derived with incorporating radiation hydrodynamics 
jointly with simple stellar evolution in a bulge. 
As a result, the BH-to-bulge ratio is basically determined 
by the energy conversion efficiency for nuclear fusion 
of hydrogen to helium, $\varepsilon=0.007$.
Also, in relation to BH growth, a scenario for quasar formation 
is addressed.

\section{Radiation Drag-Induced Mass Accretion} \label{SMBH}

We suppose a simple two-component system 
which consists of a spherical stellar bulge and dusty gas within it. 
The exact expressions for the radiation drag are found in some 
literatures (Umemura, Fukue, \& Mineshige 1997;
Fukue, Umemura, \& Mineshige 1997).
For the total luminosity $L_*$ of a uniform bulge, 
the radiation energy density is given by $E \simeq L_*/cR^2$, where
$c$ is the light speed and $R$ is the radius of the bulge.
Then, the angular momentum loss rate by the radiation drag is given by 
$
d \ln J / dt \simeq - \chi E / c,
$
where $J$ is the total angular momentum of gaseous component and
$\chi$ is the mass extinction coefficient which is given by $\chi = \kappa /\rho$
with dust absorption coefficient $\kappa$ and gas density $\rho$.
Therefore, in an optically-thin regime, 
$
d \ln J / dt \simeq - \tau L_* / c^2 M_g,
$
where $\tau$ is the total optical depth of the system
and $M_g$ is the total mass of gas.
In an optically-thick regime, the radiation drag efficiency is saturated
due to the conservation of the photon number (Tsuribe \& Umemura 1997). Thus, 
an expression of the angular momentum loss rate 
which includes both regimes is given by
$
d \ln J / dt \simeq - (L_* / c^2 M_g) (1-{\rm e}^{-\tau}).
$
In practice, it is likely that optically-thin surface layers are stripped 
from optically-thick clumpy clouds by the radiation drag, and
the stripped gas losing angular momentum accretes onto the center
(Kawakatu \& Umemura 2001).
Since the radiative cooling is effective in the surface layers, the 
accretion is likely to proceed in an isothermal fashion
until an optically-thick massive dark object forms.
Then, the mass accretion rate is estimated to be
\begin{equation}
\dot{M} = -M_g{d \ln J \over dt} \simeq {L_* \over c^2}(1-{\rm e}^{-\tau}).
\end{equation}
In an optically-thick regime, this gives simply 
$
\dot{M} = {L_* / c^2},
$
which is numerically 
$
\dot{M} = 0.1 M_\odot {\rm yr}^{-1} (L_* / 10^{12} L_\odot).
$
This rate is comparable to the Eddington mass accretion rate for
a black hole with $10^8 M_\odot$, that is,
$
\dot{M}_{\rm Edd} = 0.2 M_\odot {\rm yr}^{-1} \eta^{-1} 
(M_{\rm BH} / 10^{8} M_\odot),
$
where $\eta$ is the energy conversion efficiency. For the moment, 
$\eta=0.42$ for an extreme Kerr black hole is assumed.
For the moment, an optically-thick stage is considered.
Then, the mass of an MDO is estimated by
\begin{equation}
M_{\rm MDO} =\int^t_0 \dot{M} dt \simeq \int^t_0 L_*/c^2 dt. \label{m-mdo}
\end{equation}

The next task is to construct a model for bulge evolution.
Here, we employ a simplest analytic model. 
The gas fraction $f_g\equiv M_g/M_b$ of the system is regulated by
\begin{equation}
{ df_g \over dt} = -S(t)+F(t),
\end{equation}
where $S(t)$ is the star formation rate which is assumed to be
a Schmidt law, $kf_g$, and $F(t)$ is the mass loss rate from stars
which includes supernova explosions and stellar winds.
If we invoke the instantaneous recycling approximation, then
$F(t)=S(t)(1-\alpha)$, where $\alpha$ is the net efficiency of
the conversion into stars after subtracting the mass loss.
Then, we have an analytic solution as $f_g={\rm e}^{-\alpha kt}$.
Hence, the star formation rate is
$
\dot{M}_*/M_b=S(t)=k{\rm e}^{-\alpha kt}.
$
The radiation energy emitted by a main sequence star over its lifetime
is $0.14\varepsilon$ to the rest mass energy,
where $\varepsilon$ is the energy conversion efficiency of 
nuclear fusion from hydrogen to helium, which is 0.007.
Thus, the luminosity of the bulge is estimated to be
$
L_*=0.14\varepsilon k{\rm e}^{-\alpha kt}M_b c^2
$.
By substituting this in (\ref{m-mdo}), we find
\begin{equation}
M_{\rm MDO}=0.14\varepsilon \alpha^{-1}M_b(1-{\rm e}^{-\alpha kt}). 
\label{m-mdo2}
\end{equation}
The term $M_b(1-{\rm e}^{-\alpha kt})$ represents just the stellar 
mass in the system, that is, $M_{\rm bulge}$ observationally.
As a result, the MDO mass to bulge mass ratio is given by
\begin{equation}
M_{\rm MDO}/M_{\rm bulge}=0.3\varepsilon \alpha_{0.5}^{-1},
\end{equation}
where $\alpha_{0.5}=\alpha/0.5$.
It should be noted that the ratio is time-independent and 
also regardless of the star formation rate $k$.
This result implies that the ratio is basically determined by 
the total number of emitted photons. 

\section{BH Growth and Quasar Formation} \label{QSO}

From an observational point of view, the mass of an MDO 
is often regarded as BH mass.
But, in the present model they should be distinguished from each other 
if one strictly defines the BH mass by the mass inside the event horizon. 
In the present radiation drag model, 
it is not very likely that all the angular momentum 
of stripped gas is removed thoroughly. Hence,
the MDO presumably forms a massive and compact rotating disk
due to a little residual angular momentum.
Then, the BH is more likely to grow via accretion in the MDO, 
rather than the prompt collapse of the MDO.
If the mass accretion within the MDO is driven by the viscosity, 
the AGN activity may be ignited. 
Supposing the mass accretion onto the BH horizon
is limited by an order of the Eddington rate,
the BH mass grows according to 
\begin{equation}
M_{\rm BH}=M_0 {\rm e}^{\nu t/t_{\rm Edd}} \label{m-bh},
\end{equation}
where $\nu$ is the ratio of the BH accretion rate to
the Eddington rate, $\nu=\dot{M}_{\rm BH}/\dot{M}_{\rm Edd}$, and
$t_{\rm Edd}$ is the Eddington time-scale, 
$t_{\rm Edd}=1.9 \times 10^8{\rm yr}$.
Here $M_0$ is the mass of a seed BH, which could be
a remnant of a massive population III star with 
$10^{2-3} M_\odot$ (Carr, Bond, \& Arnett 1984; Nakamura \& Umemura 2001),
or an early formed massive BH with $\sim 10^{5} M_\odot$  
(Umemura, Loeb, \& Turner 1993).
When the BH growth is given by (\ref{m-bh}),
it is delayed from the growth of the MDO given by (\ref{m-mdo2}).
In Figure 1, this delay is schematically shown.
The BH mass reaches $M_{\rm MDO}$ at a time 
$t_{\rm cross}$ when (\ref{m-bh}) equals (\ref{m-mdo2}). 
We can calculate $t_{\rm cross}$ to find 
$
t_{\rm cross} \approx 10\nu^{-1} t_{\rm Edd} \approx 10^{9}\nu^{-1} {\rm yr},
$
although it has weak dependence on $k$. As seen in Figure 1,
during $t<t_{\rm cross}$, the BH mass fraction $f_{\rm BH}$ increases with time
and therefore possibly increases with the metallicity of the system.
At $t>t_{\rm cross}$, almost all of the MDO matter has fallen onto
the central BH, and therefore the BH mass fraction is settled to
$
f_{\rm BH}= M_{\rm MDO}/M_{\rm bulge}
=0.3\varepsilon \alpha_{0.5}^{-1}=0.002\alpha_{0.5}^{-1}.
$

So far, the optically-thick regime has been assumed.
However, before $t_{\rm cross}$ the interstellar gas may be blown
out by a galactic wind and the system could be transparent.
Recently, it is argued that the color-magnitude
relation of bulges can be reproduced if a galactic wind sweeps away the gas
at the epoch of a few $10^8$yr (Kodama \& Arimoto 1997).
If the system becomes optically-thin at a wind epoch $t_{\rm w}$, 
the radiation drag-induced mass accretion
practically stops owing to the reduced efficiency of radiation
drag due to the small optical depth. 
Then, the final MDO mass is settled to
$
M_{\rm MDO}=0.14\varepsilon \alpha^{-1}M_b(1-{\rm e}^{-\alpha kt_{\rm w}}).
$ 
If the timescale of star formation $(\alpha k)^{-1}$ is longer than 
$t_{\rm w}$, the final MDO mass can be reduced.
But, $M_{\rm MDO}/M_{\rm bulge}$ is time-independent as stated above.
When all gas of the MDO eventually falls onto the BH, again
$
f_{\rm BH}=0.3\varepsilon \alpha_{0.5}^{-1}
$
is achieved at $t_{\rm cross}$.

\begin{figure}
\plotone{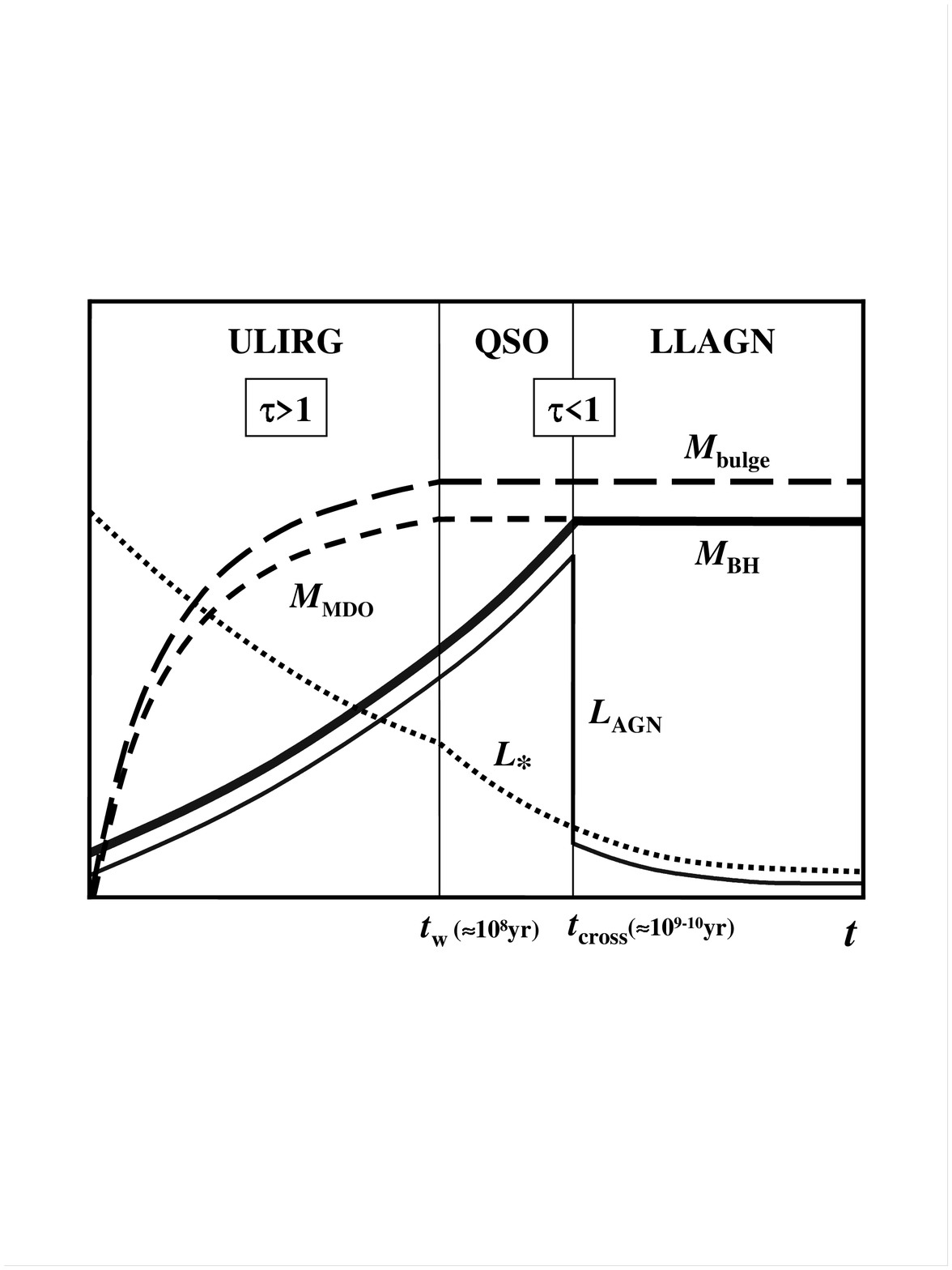}
\caption{A schematic sketch of black hole growth
and a scenario for quasar formation. 
The abscissa is time and the ordinate is mass or luminosity
in arbitrary units. 
$M_{\rm bulge}$ is the mass of stellar component in the bulge.
$M_{\rm MDO}$ is the mass of the massive dark object (MDO).
$M_{\rm BH}$ is the mass of the supermassive BH.
$L_*$ and $L_{\rm AGN}$ are the bulge luminosity and
the black hole accretion luminosity, respectively. 
$t_{\rm w}$ is the galactic wind timescale, and
$t_{\rm cross}$ is defined so that $M_{\rm MDO}=M_{\rm BH}$.
\label{fig1}}
\end{figure}

Based on the present model for BH growth,
a picture of quasar formation may be constructed as 
shown schematically in Figure 1.
Recently, the Eddington ratio is estimated 
to be $\nu \approx 0.1$ for quasars 
(McLeod, Rieke, \& Storrie-Lombardi 1999).
Hence, $t_{\rm cross} \approx 10^{10}{\rm yr}$
and therefore $t_{\rm cross} \gg t_{\rm w}$.
Since the bulge luminosity decreases as $L_* \propto {\rm e}^{-\alpha kt}$,
the stage at $t < t_{\rm w}$ is a bright, optically-thick phase,
which may correspond to a ultraluminous infrared galaxy (ULIRG).
Then, a ULIRG ($t<t_{\rm w}$) harbors a more or less active nucleus.
This can be regarded as a model for a paradigm of the evolution of ULIRGs 
to quasars proposed by Sanders et al. (1998) and Norman \& Scoville (1998).
Even at $t>t_{\rm w}$, 
$M_{\rm BH}$ still continues to grow until $t_{\rm cross}$,
and therefore the AGN brightens with time if the Eddington ratio is constant.
This optically-thin phase may correspond to quasar phenomena.
Its early phase might be observed as a narrow line type 1 AGN (Mathur 2001).
After the AGN luminosity ($L_{\rm AGN}$) exhibits 
a peak at $t_{\rm cross}$, it fades out abruptly, 
not only because the mass accretion rate drops promptly
in the optically-thin phase, but also
because the energy conversion efficiency is in proportion to
$\dot{M}/\dot{M}_{\rm Edd}$ for an optically-thin
advection-dominated accretion flow
(e.g. Kato, Fukue, \& Mineshige 1998).
The later fading nucleus could be a low luminosity AGN (LLAGN)
(e.g. Kawaguchi \& Aoki 2001). 
As a consequence, $f_{\rm BH}$ increases with increasing
$L_{\rm AGN}$ up to
$
f_{\rm BH}=0.3\varepsilon \alpha_{0.5}^{-1}.
$
Then, $f_{\rm BH}$ for quasars is predicted to be similar to
that of ellipticals, although possibly a bit smaller.
As for $f_{\rm BH}$ for Seyfert 1 galaxies, naively it is expected to
be level with that for quasars, but 
circumnuclear starbursts might predominantly regulate 
the mass accretion (Umemura, Fukue, \& Mineshige 1998).

Finally, we point out that
further mass accretion could be induced also by the AGN 
luminosity.
The mass accumulated by this self-induced accretion is maximally
$
M_{\rm Ind}=\eta (1-\eta)^{-1}M_{\rm BH},
$
if the successive induction is included.
Hence, at later stages as $t>t_{\rm cross}$, we have
$
f_{\rm BH}=(M_{\rm MDO}+M_{\rm Ind})/M_{\rm bulge}.
$
Thus, the BH-to-bulge mass ratio could become
\begin{equation}
f_{\rm BH}=0.14\varepsilon \alpha^{-1}(1-\eta)^{-1}
=0.5\varepsilon \alpha_{0.5}^{-1}, \label{m-ratio}
\end{equation}
if an extreme Kerr BH ($\eta =0.42$) is assumed, whereas 
the self-induction factor $(1-\eta)^{-1}=1.06$ 
for a Schwartzschild BH ($\eta =0.057$) brings
no substantial increase on $ f_{\rm BH}$.
In the long run, although leaving some uncertainties with 
evolution model including $\alpha_{0.5}$, 
$f_{\rm BH}$ is predicted to be $0.3\varepsilon -0.5\varepsilon$.

\section{Discussion} \label{Discussion}

The radiation drag efficiency could be strongly subject to the 
effect of geometrical dilution (Umemura, Fukue, \& Mineshige 1998;
Ohsuga et al. 1999).
If the system is spherical, the emitted photons are effectively consumed 
within the system, whereas a large fraction of photons can escape 
from a disk-like system and thus the drag efficiency tends to be 
considerably reduced. Although this is very qualitative,
the geometrical dilution may be the reason why $f_{\rm BH}$ is
observed to be significantly smaller than 0.001 in disk galaxies.
For the same reason, $f_{\rm BH}$ could be lower for a flattened bulge.

Finally, the present mechanism may provide a physical basis for the 
state-of-the-art cosmological scenarios for the joint formation of galaxies
and quasars (Haehnelt \& Rees 1993; Haiman \& Loeb 1998;
Kauffmann \& Haehnelt 2000; Monaco, Salucci, \& Danese 2000; 
Granato et al. 2001; Hosokawa et al. 2001).
In cosmological studies, $f_{\rm BH}$ has been hitherto parameterized 
so that it should match to the observed BH-to-bulge mass relation. 
Also, Monaco, Salucci, \& Danese (2000) argue that, in order to account
for the observed statistics of quasars and elliptical galaxies, there
should be a time delay between the beginning of the star formation
and the quasar bright phase. In Granato et al. (2001), the formation
of early-type protogalaxies as well as quasars is successfully accounted
for if it is assumed that the relationship between the BH mass and 
the host mass has been imprinted during the early phase of the quasar and host
evolution. In the present paper, the BH-to-bulge mass relation
is a function of time, but it has been found that 
in a quasar phase their assumption holds good. 
In the light of these points, 
the basic assumptions and results by previous 
cosmological scenarios for quasar formation are mostly consistent
with the present model for quasar formation based on the BH growth.

\acknowledgments

The author thanks A. Ferrara, N. Kawakatu, S. Mineshige, 
T. Nakamoto, K. Ohsuga, and H. Susa
for fruitful discussions. Also, the author is indebted to the support by
Center for Computational Physics, University of Tsukuba.
This work was supported in part by the Grant-in-Aid of the JSPS, 11640225.

\newpage
\begin{center}
{\bf REFERENCES}
\end{center}
\re	Adams, F. C., Graff, D. S., \& Richstone, D. O. 2001, ApJ, 551, L31
\re	Bahcall, J. N., et al. 1997, ApJ, 479, 642
\re	Brotherton, M. S., et al. 1999, ApJ, 520, L87
\re	Carr, B. J., Bond, J. R., \& Arnett, W. D. 1984, ApJ, 277, 445
\re	Ferrarese, L., \& Merritt, D. 2000, ApJ, 539, L9
\re	Fukue, J., Umemura, M., \& Mineshige, S. 1997, PASJ, 49, 673
\re	Gebhardt, K., et al. 2000a, ApJ, 539, L13
\re	------------. 2000b, ApJ, 543, L5
\re	Granato, G. L., Silva, L., Monaco, P., Panuzzo, P., Salucci, P., 
	Zotti, G. De, \& Danse, L. 2001, MNRAS, 324, 757
\re	Haehnelt, M. G., \& Rees, M. J. 1993, MNRAS, 263, 168
\re	Haiman, Z. \& Loeb, A. 1998, ApJ, 503, 505
\re	Hooper, E. J., Impey, C. D., \& Foltz, C. B. 1997, ApJ, 480, L95
\re	Hosokawa, T., Mineshige, S., Kawaguchi, T. Yoshikawa, K. 
	\& Umemura, M. 2001, PASJ, in press
\re	Kato, S., Fukue, J., \& Mineshige, S. 1998, Black-Hole Accretion Disk
	(Kyoto: Kyoto Univ. Press)
\re	Kauffmann, G.,  \& Haehnelt, M. 2000, MNRAS, 311, 576
\re	Kawaguchi, T., \& Aoki, K. 2001, preprint
\re	Kawakatu, N., \& Umemura, M. 2001, MNRAS, submitted.
\re	Kirhakos, S., Bahcall, J. N., Schneider, D. P., \& Kristian, J.
	1999, ApJ, 520, 67
\re	Kodama, T., \& Arimoto, N. 1997, A\&A, 320, 41
\re	Kormendy, J., \& Richstone, D. 1995, ARA\&A, 33, 581
\re	Krolik, J. H. 2001, ApJ, 551, 72
\re	Laor, A. 1998, ApJ, 505, L83
\re	------------. 2001, ApJ, 553, 677
\re	Magorrian, J., et al. 1998, AJ, 115, 2285
\re	Mathur, S. 2001, AJ, in press (astro-ph/0107163)
\re	McLeod, K. K., \& Rieke, G. H. 1995, ApJ, 454, L77
\re	McLeod, K. K., Rieke, G. H. \& Storrie-Lombardi, L. J. 1999, ApJ, 511, L67
\re	McLure, R. J., Dunlop, J. S., \& Kukula, M. J. 2000, MNRAS, 318, 693
\re	McLure, R. J., Kukula, M. J., Dunlop, J. S., Baum, S. A.,
	O'Dea, C. P., \& Hughes, D. H. 1999, MNRAS, 308, 377
\re	McLure, R. J., \& Dunlop, J. S. 2001, MNRAS, in press 
	(astro-ph/0009406) 
\re	Merrifield, M. R., Forbes, Duncan A., \& Terlevich, A. I. 2000, 
		MNRAS, 313, L29
\re	Merritt, D., \& Ferrarese, L. 2001a, MNRAS, 320, L30
\re	------------. 2001b, ApJ, 547, 140
\re	Mineshige, S., Tsuribe, T., \& Umemura, M. 1998, PASJ, 50, 233
\re	Monaco, P., Salucci, P., \& Danese, L. 2000, MNRAS, 311, 279
\re	Nakamura, F., \& Umemura, M. 2001, ApJ, 548, 19
\re	Nelson, C. H. 2000, ApJ, 544, L91
\re	Norman, C. \& Scoville, N. 1988, ApJ, 332, 124
\re	Ostriker, J. P. 2000, Phys. Rev. Lett., 84, 5258
\re	Ohsuga, K., \& Umemura, M. 2001, ApJ, in press (astro-ph/0105474) 
\re	Ohsuga, K., Umemura, M., Fukue, J., \& Mineshige, S. 1999, PASJ, 51, 345 
\re	Rees, M. 1984, ARA\&A, 22, 471
\re	Richstone, D., et al. 1998, Nature, 395A, 14
\re	Salucci, P., et al. 2000, MNRAS, 317, 488
\re	Sarzi, M., et al. 2001, ApJ, 550, 65
\re	Sasaki, S. \& Umemura, M. 1996, ApJ, 462, 104
\re	Silk, J., \& Rees, M. 1998, A\&A, 331, L1
\re	Tsuribe, T., \& Umemura, M. 1997, ApJ, 486, 48
\re	Umemura, M., Fukue, J., \& Mineshige, S. 1997, ApJ, 479, L97
\re	------------. 1998, MNRAS, 299, 1123
\re	Umemura, M., Loeb, A., \& Turner, E. L. 1993, 419, 459
\re	Wandel, A. 1999, ApJ, 519, L39

\end{document}